
\input phyzzx
\overfullrule=0pt

\def\gl{\lambda}
\def\p{\partial}
\def\phip{\p_+\phi}
\def\phim{\p_-\phi}

\def\rhop{\p_+\rho}
\def\rhom{\p_-\rho}
\def\rhopm{\p_+\p_-\rho}
\def\chip{\p_+\chi}
\def\chim{\p_-\chi}
\def\chipm{\p_+\p_-\chi}
\def\omp{\p_+\Omega}
\def\omm{\p_-\Omega}
\def\ompm{\p_+\p_-\Omega}
\def\hxp{\hat x^+}
\def\hxm{\hat x^-}

\rightline {UTTG-19-92}
\rightline {SU-ITP-92-24}
\rightline {August 1992}
\bigskip\bigskip
\title{Cosmic Censorship in Two-Dimensional Gravity\foot{Work
supported in part by NSF grants PHY89-17438, PHY 9009850 and
R.~A.~Welch Foundation.}}

\vfill
\author{Jorge G. Russo,\foot{INFN associate, e-mail:
russo@utaphy.bitnet} }
\address {Theory Group, Department of Physics, University of
Texas\break
Austin, TX 78712}
\author{Leonard Susskind and L\'arus
Thorlacius\foot{e-mail: larus@dormouse.stanford.edu}}
\address{ Department of Physics \break Stanford University,
            Stanford, CA 94305}
\vfill

\abstract
\singlespace
A weak version of the cosmic censorship hypothesis is implemented as
a set of boundary conditions on exact semi-classical solutions of
two-dimensional
dilaton gravity.  These boundary conditions reflect low-energy matter
from the strong coupling region and they also serve to stabilize the
vacuum of the theory against decay into negative energy states.
Information about low-energy incoming matter can be recovered in
the final state but at high energy black holes are formed and
inevitably lead to information loss at the semi-classical level.

\vfill\endpage

\REF\hawi{S.~W.~ Hawking
\journal Comm .Math. Phys. & 43 (75) 199.}

\REF\hawii{S.~W.~Hawking
\journal Phys. Rev. & D14 (76) 2460.}

\REF\cghs{C.~G.~Callan, S.~B.~Giddings, J.~A.~Harvey and
A.~Strominger
\journal Phys. Rev. & D45 (92) R1005.}

\REF\bddo{T.~Banks, A.~Dabholkar, M.~R.~Douglas and M.~O'Loughlin
\journal Phys. Rev. & D45 (92) 3607.}

\REF\rst{J.~G.~Russo, L.~Susskind and L.~Thorlacius, {\it Black
Hole Evaporation in 1+1 Dimensions}, Stanford University preprint,
SU-ITP-92-4, January 1992, to appear in Physics Letters {\bf B}.}

\REF\bghs{B.~Birnir, S.~B.~Giddings, J.~A.~Harvey and A.~Strominger
\journal Phys. Rev. & D46 (92) 638.}

\REF\haw{S.~W.~Hawking
\journal Phys. Rev. Lett. & 69 (92) 406.}

\REF\lslt{L.~Susskind and L.~Thorlacius, {\it Hawking Radiation and
Back-Reaction}, SU-ITP-92-12, hepth@xxx/9203054, March 1992,
to appear in Nucl. Phys.~{\bf B}.}

\REF\strom{A.~Strominger, {\it Fadeev-Popov Ghosts and 1+1
Dimensional Black
Hole Evaporation}, UC Santa Barbara preprint, UCSB-TH-92-18,
hepth@xxx/9205028,
May 1992.}

\REF\jrat{J.~G.~Russo and A.~A.~Tseytlin
\journal Nucl. Phys. & B382 (92) 259.}

\REF\chams{T.~Burwick, A.~Chamseddine, to appear in Nucl. Phys. {\bf B}.}

\REF\bilcal{A.~Bilal and C.~G.~Callan, {\it Liouville Models of Black
Hole
Evaporation}, Princeton University preprint, PUPT-1320,
hepth@xxx/9205089, May
1992.}

\REF\dealw{S.~P.~de~Alwis, {\it Quantization of a Theory of 2d
Dilaton
Gravity}, University of Colorado preprint, COLO-HEP-280,
hepth@xxx/9205069, May
1992; {\it Black Hole Physics from Liouville Theory}, COLO-HEP-284,
hepth@xxx/9206020, June 1992; {\it Quantum Black Holes in Two
Dimensions}, COLO-HEP-288, hepth@xxx/9207095, July 1992.}

\REF\endpoint{J.~G.~Russo, L.~Susskind and L.~Thorlacius, {\it The
Endpoint of Hawking Radiation}, Stanford University preprint,
SU-ITP-92-17, June 1992, to appear in Physical Review {\bf D}.}

\REF\gidstr{S.~B.~Giddings and A.~Strominger, {\it Quantum Theories
of Dilaton Gravity}, UC Santa Barbara preprint, UCSB-TH-92-28,
hepth@xxx/9207034, July 1992.}

\REF\hawste{S.~W.~Hawking and J.~M.~Stewart, {\it Naked and
Thunderbolt Singularities in Black Hole Evaporation}, University of
Cambridge preprint, hepth@xxx/9207105, July 1992.}

\REF\chrful{S.~M.~Christensen and S.~A.~Fulling
\journal Phys. Rev. & D15 (77) 2088.}

\REF\bjorn{B.~Birnir and A.~Strominger, unpublished.}

\REF\penrose{R.~Penrose
\journal Rivista del Nuovo Cimento & 1 (69) 252.}

\REF\witten{E.~Witten
\journal Phys. Rev. & D44 (91) 314.}

\REF\bps{T.~Banks, M.~Peskin and L.~Susskind
\journal Nucl. Phys. & B244 (84) 125.}

\chapter{Introduction}

Black hole physics provides a setting for the study of the
interplay between general relativity and quantum mechanics.  In
particular, it appears difficult to reconcile the apparently thermal
evaporation of a black hole formed in gravitational collapse with the
hamiltonian evolution of pure quantum mechanical states
[\hawi,\hawii].  In recent months considerable effort has been put
into developing a
semi-classical description of black hole evolution in two-dimensional
dilaton gravity coupled to conformal matter [\cghs-\hawste].  This
simplified context shares important features with more realistic
four-dimensional black hole physics.  The original model proposed by
Callan, Giddings, Harvey and Strominger [\cghs] (CGHS) has singular
classical solutions, which describe the formation of a black hole by
incoming matter, and the Hawking emission from this background
geometry can be obtained from the conformal anomaly of the matter
fields [\chrful,\cghs].  CGHS further suggested a semi-classical
description of the back-reaction of the Hawking radiation on the
geometry by introducing anomaly-induced terms into the equations of
motion of the model.

The quantum corrected CGHS-equations have not been solved in closed
form\foot{Numerical results on black hole evolution in the CGHS-model
have been obtained in [\lslt,\hawste,\bjorn].} but soon after the
original work of CGHS it was shown that gravitational collapse always
leads to a curvature singularity at a certain critical value of the
dilaton field in their theory [\bddo,\rst].  Modifications of
the model have since been found, where the formation of a singularity
in collapse can be avoided [\strom].  Furthermore, an improved
treatment of the quantum theory [\jrat-\gidstr] has lead to models
where the semi-classical equations can be solved exactly, as first
exhibited by Bilal and Callan [\bilcal] and de~Alwis [\dealw].

In this paper we will argue that having a critical value of the
dilaton, where a singularity forms in collapse, is really a blessing
in disguise, and may prove essential for a consistent formulation of
a quantum theory of two-dimensional black holes.  The possibility
that no singularity forms in gravitational collapse is appealing but
unfortunately all solvable two-dimensional models suggested to date
which exhibit this feature suffer from serious instabilities
[\bilcal,\dealw,\gidstr].

In a previous paper [\endpoint] we proposed a particular solvable
model which has some desirable features built into it.
The linear dilaton vacuum is a solution. Furthermore,
all static solutions with negative ADM-energy have naked
singularities.  We were able to follow the evolution of a black hole,
formed by the gravitational collapse of a shock wave, using an exact
solution of the semi-classical equations.  As expected, the
collapsing matter forms a space-like singularity at a critical value
of the dilaton field inside an apparent horizon.  As the black hole
evaporates the apparent horizon recedes and after a finite proper
time it meets the singularity.  At that point the singularity is no
longer cloaked by the horizon and the future evolution of the
geometry is not uniquely determined.  In [\endpoint] we showed that
there exist boundary
conditions which match the final state of the black hole evolution
onto the linear dilaton vacuum.  With this choice of boundary
conditions the geometry is non-singular after the black hole
evaporation is complete.

Here we will study this model further.
A two-dimensional version of the cosmic censorship hypothesis
[\penrose] will play a central role in our considerations.  At the
endpoint of Hawking evaporation of a black hole, when the apparent
horizon meets the singularity, a region of strong curvature becomes
visible to outside observers.  This constitutes a violation of the
cosmic censorship hypothesis, but a fairly mild one, with the naked
singularity being an isolated event.\foot{We thank A.~Strominger for
discussions on this point.}  If we choose other boundary conditions,
which do not match the black hole solution onto the vacuum when the
evaporation is completed, the black hole singularity becomes
time-like at that point and persists in the future geometry.  We
will, however, insist that physical configurations do not develop
such extended naked singularities and use this requirement to
determine physical boundary conditions in the strong coupling region.
 In other words, we will implement a weak form of the cosmic
censorship hypothesis which states that curvature singularities in
our semi-classical geometries must be hidden behind an apparent
horizon, except for isolated events such as the endpoint of Hawking
evaporation of a black hole.

We will take the line $\phi =\phi_{\rm cr}$ to be a
boundary of spacetime in the semi-classical theory.\foot{It is useful
to keep in mind an analogy
with the dimensional reduction to radial degrees
of freedom of higher dimensional gravity.  In the effective
two-dimensional theory spacetime has a boundary at the origin of the
radial coordinate.}  In `low-energy physics$ $' where the incoming
energy flux is below a certain threshold value (which equals the rate
at which a black hole evaporates\foot{The evaporation rate of
two-dimensional black holes is independent of their mass
[\witten,\cghs]}) no black holes are formed in scattering processes
[\endpoint].  In this case the critical line remains time-like and we
have to supply boundary conditions for the fields there.
We will show how requiring the curvature to be finite at $\phi
=\phi_{\rm cr}$, or in other words not allowing a naked singularity
there, leads to
the reflecting boundary conditions we suggested in our
previous paper.  The fact that such non-singular boundary conditions
can be found, and their physical interpretation in terms of
reflecting energy, support our view that the spacetime has a boundary
at $\phi =\phi_{\rm cr}$.  This time-like boundary is not present in
the classical theory and as a result incoming matter cannot be turned
away classically from the strong coupling region regardless of how
little
energy it
carries.  The quantum corrections spontaneously generate a boundary
so that low energy matter is reflected in the semi-classical
theory and a black hole is only formed if the incoming matter carries
a certain minimum energy density [\endpoint].  At the same time the
boundary conditions stabilize the theory against decay
into negative energy configurations by implementing cosmic censorship
at the critical line $\phi =\phi_{\rm cr}$.

In the following section we write down some key definitions in order
to establish notation and make contact with previous work.
We also point out that the functional integral over all the
fields can be carried out explicitly in some models where no
spacetime
boundary is present.  Such theories thus
appear too simple to describe the physics of two-dimensional quantum
gravity.  However, the quantum theory becomes non-trivial with
boundaries present, especially since in this case not only the
two-dimensional spacetime but also the field space of the theory is
bounded at the semi-classical level.  In section~3 we consider
semi-classical solutions
describing general incoming matter distributions with energy flux
below the threshold required to form black holes.  We use cosmic
censorship to derive boundary conditions at $\phi =\phi_{\rm cr}$ for
this low energy physics and then we verify that these boundary
conditions are consistent with energy conservation.

\chapter{Semi-classical model}

We will work with the two-dimensional dilaton gravity model we
introduced in [\endpoint].  It is related to
models studied by Bilal and Callan [\bilcal] and de~Alwis
[\dealw] and reduces to the original CGHS-theory at the classical
level.  The semi-classical equations are derived from the one-loop
effective action
$$\eqalign{
S = {1\over \pi} \int d^2x\>
\bigl[e^{-2\phi}(2&\rhopm - 4 \phip \phim
+\gl ^2 e^{2\rho})
+{1\over 2}\sum_{i=1}^N \p_+ f_i \p_- f_i  \cr
&- {\kappa\over \pi}
(\rhop\rhom +\phi\rhopm) \bigr]\> ,  \cr}
\eqn\first
$$
written here in conformal gauge, $g_{++}=g_{--}=0$, $g_{+-}=-{1\over
2}
e^{2\rho}$.  If the coefficient $\kappa$ in front of the one-loop
quantum correction terms has the value $\kappa = {N-24\over 12}$ this
action defines a conformal field theory [\jrat-\dealw].
In everything that follows we will assume
that the number of matter fields is $N>24$, so that $\kappa$
is positive.  The first quantum correction term in \first\ comes from
the one-loop conformal anomaly and the second one is a covariant,
local
counterterm which we are free to add to the definition of our model.
The classical theory has a symmetry generated by the current
$$
j^\mu=\partial ^\mu (\phi - \rho)  \> .
\eqn\second
$$
The exactly soluble semi-classical models all have a corresponding
symmetry but if our counterterm is added the current maintains the
same simple form in terms of $\rho $ and $\phi$ at the semi-classical
level.  By defining our model this way we simplify the analysis and
interpretation of the semi-classical solutions considerably.  For
example, the linear dilaton solution of the classical theory is
preserved in our model.

In addition to the equations derived from \first\ we have to impose
as constraints the equations of motion of the metric components which
are set to zero in this gauge,
$$\eqalign{
0=T_{\pm\pm}= (e^{-2\phi}-{\kappa\over 4})\,
(4 &\p_\pm\rho \p_\pm\phi - 2\p_\pm^2\phi)
 + {1\over 2} \sum_{i=1}^N \p_\pm f_i \p_\pm f_i \cr
 &- \kappa (\p_\pm\rho \p_\pm\rho - \p_\pm^2\rho
     +t_\pm) \> . \cr}
\eqn\third
$$
The non-local character of the conformal anomaly is expressed in the
functions $t_\pm(x^\pm)$ which are to be fixed by physical boundary
conditions on the matter energy momentum tensor.

The action simplifies dramatically if we make the following field
redefinitions,
$$\eqalign{
\Omega =& {\sqrt{\kappa}\over 2} \phi
+ {e^{-2\phi}\over \sqrt{\kappa}}\> , \cr
\chi =& \sqrt{\kappa} \rho - {\sqrt{\kappa}\over 2} \phi
+ {e^{-2\phi}\over \sqrt{\kappa}}\> . \cr}
\eqn\fourth
$$
In terms of the new variables the action \first\ becomes
$$
S = {1\over \pi} \int d^2x\>
\bigl[-\chip\chim + \omp\omm
+ \gl^2 e^{{2\over \sqrt{\kappa}}(\chi -\Omega)}
+{1\over 2}\sum_{i=1}^N \p_+ f_i \p_- f_i \bigr]\> ,
\eqn\fifth
$$
and the constraints \third\ reduce to
$$
\kappa \, t_\pm = -\p_\pm\chi \p_\pm\chi
+\sqrt{\kappa}\, \p_\pm^2\chi
+\p_\pm\Omega \p_\pm\Omega
+ {1\over 2} \sum_{i=1}^N \p_\pm f_i \p_\pm f_i \> .
\eqn\sixth
$$
In the following section we will consider semi-classical solutions of
this system.

We conclude this section with some observations about the quantum
theory defined by the action \fifth .  The apparent similarity to
Liouville
theory is somewhat deceptive
because the path integral over $\Omega$ and $\chi$ can in fact be
carried out explicitly, leading to an effective action which turns
out to be identical to the classical action \fifth .  To see this we
first rewrite the action in terms of `light-cone$ $'
variables in field space,
$$
S[\Omega_+,\Omega_-]= {1\over \pi} \int d^2x\>
\bigl[-\p_+\Omega_+\p_-\Omega_-
+ \gl^2 e^{{2\over \sqrt{\kappa}}\Omega_-} \bigr]\> ,
\eqn\seventh
$$
where $\Omega_\pm = \chi \pm \Omega$ and we have dropped the matter
term from the action for the time being.  Now we couple the new
fields to sources and consider the generating functional,
$$
e^{i W[j^+,j^-]} = {\cal N}
\int [d\Omega_+][d\Omega_-]\,
e^{iS[\Omega_+,\Omega_-] +i\int d^2x\,(j^+\Omega_+ + j^-\Omega_-)}
\>,
\eqn\eighth
$$
where $\cal N$ is a normalization constant.
Only $\Omega_-$ appears in the interaction term in the action
\seventh\ so we can carry out the functional integral over
$\Omega_+$.  The result is a formal delta-function involving
$\Omega_-$ which means that we can also perform the second functional
integral in \eighth .  The connected generating functional is
$$
W[j^+,j^-]= \int d^2x\>
\bigl[{\gl^2\over \pi}e^{-{2\pi\over \sqrt{\kappa}}
\int d^2x '\> g(x,x ') j^+(x ')}
-\pi \int d^2x '\> j^-(x) g(x,x ') j^+(x ')\bigr] \> ,
\eqn\ninth
$$
where the Green's function is a solution of
${\p\over \p x^+}{\p\over \p x^-}g(x,x ')=\delta^{(2)}(x-x ')$.
The final step is to obtain the effective action via a Legendre
transform,
$$\eqalign{
\Gamma[\Omega_+^{\rm cl},\Omega_-^{\rm cl}]=&
W[j^+,j^-]- \int d^2x\>
(j^+\Omega_+^{\rm cl}+j^-\Omega_-^{\rm cl}) \cr
=&{1\over \pi} \int d^2x\>
\bigl[-\p_+\Omega_+^{\rm cl}\p_-\Omega_-^{\rm cl}
+ \gl^2 e^{{2\over \sqrt{\kappa}}\Omega_-^{\rm cl}}
\bigr]\> ,  \cr}
\eqn\tenth
$$
where we have used that
$\Omega_-^{\rm cl}(x)={\delta W\over \delta j^-(x)}
=-\pi\int d^2x '\> g(x,x ') j^+(x ')$.

This formal argument is of course not sufficient by itself to prove
that the action \fifth\ receives no quantum corrections.  We have to
introduce regularization, choose boundary conditions on propagators
\etc\ to make these expressions well
defined.  The point, however, is simply that the kinetic terms of
$\Omega$ and $\chi$ in \fifth\ have opposite signs while these fields
appear in a symmetric fashion in the interaction term.  As a result
one finds a lot of cancellation in a diagrammatic perturbation
expansion.  If the $\Omega$ and $\chi$ propagators are defined using
identical regularization and boundary conditions they will differ
only by a
sign and any two diagrams which differ only by a single internal
propagator will exactly cancel.  This means in particular that all
loop diagrams will cancel one-on-one and the full effective action
will be generated by tree graphs as suggested by \tenth .

A key assumption in the above argument is that the $\Omega$ and
$\chi$ propagators satisfy the same boundary conditions.  In our
semi-classical theory the critical line $\phi=\phi_{\rm cr}$ is a
spacetime boundary.  Quantum consistency conditions may require
introducing non-trivial boundary interactions there, or even new
degrees of freedom, and these will in general not preserve the
symmetry between $\Omega$ and $\chi$.  Note also that $\Omega$ in
\fourth\ is bounded from below in our semi-classical theory.  It
takes its minimum value at $\phi=\phi_{\rm cr}$.  Restricting the
range of $\Omega$ in the quantum path integral leads to non-trivial
physical effects.  Away from the boundary quantum fluctuations can
cause the
dilaton field to reach its critical value in some region which is
then no longer part of the spacetime.  The full quantum theory will
then include configurations with disconnected boundary
components in the path integral.  Topology change of this kind would
be strongly suppressed in the weak coupling region where asymptotic
observers are located but it may play an important role in the strong
coupling physics near the boundary.

In some
models with $N<24$ matter fields the singularity at $\phi=\phi_{\rm
cr}$ is absent and the issue of restricting the range of $\Omega$
does not arise [\strom,\bilcal,\dealw].  In this case boundary
conditions are imposed on propagators in the asymptotic regions of
weak and strong coupling.  If they are chosen so as to preserve
symmetry between $\Omega$ and $\chi$ the quantum theory will be very
simple indeed, as we argued above.

\chapter{Low energy physics, boundary conditions and cosmic
censorship}

The semi-classical equations of motion derived from the effective
action \fifth\ are
$$
\chipm = \ompm =
-{\gl^2\over \sqrt{\kappa}}
e^{{2\over \sqrt{\kappa}}(\chi-\Omega)} \> .
\eqn\eleventh
$$
This set of equations can be explicitly integrated [\bilcal,\dealw]
and we will focus our attention on solutions which describe the
response to a general distribution of incident matter.  It is easily
checked that the linear dilaton vacuum, which takes the form
$e^{-2\phi}=e^{-2\rho}=-\gl^2 x^+x^-$ in `Kruskal$ $' coordinates,
solves the equations of motion and we will match solutions with
incident matter onto this vacuum in the far past.

The classical energy flux carried by an arbitrary distribution of
incoming matter is described by some $T^f_{++}(x^+)={1\over
2}\p_+f\p_+f$ at $x^-\rightarrow -\infty$.  For convenience we will
assume that the matter arrives over a finite span of time so that
$T^f_{++}(x^+)$ vanishes for $x^+<x^+_0$ and $x^+>x^+_1$ (see
figure~1).  The limits $x^+_0\rightarrow 0$ and
$x^+_1\rightarrow\infty$ can be taken at the end of the day if
desired.

At a given value of $x^+$ one can define the integrated incoming
energy and Kruskal momentum up to that point,
$$\eqalign{
M(x^+) =& \gl \int_0^{x^+} dx^+ \,
x^+ \, T^f_{++}(x^+) \> ,
\cr
P_+(x^+) =& \int_0^{x^+} dx^+ \, T^f_{++}(x^+) \> ,
\cr}
\eqn\twelfth
$$
and it turns out that the incoming energy flux only enters into the
semi-classical solution through these two functions.  It is easy to
verify that the equations of motion \eleventh\ and the constraints
\sixth\ are satisfied by
$$
\Omega =\chi =
- {\gl^2\over \sqrt{\kappa}}\,
x^+\bigl(x^-+{1\over \gl^2}P_+(x^+)\bigr)
+ {M(x^+)\over \sqrt{\kappa}\gl}
-{\sqrt{\kappa}\over 4}\log{(-\gl^2 x^+x^-)} \> .
\eqn\thirteenth
$$
This solution is valid for all $x^-<x^-_0$ (region (i) in figure 1)
and it reduces to the linear dilaton vacuum for $x^+<x^+_0$.

At $x^\pm =x^\pm_0$ the leading edge of the incoming matter
distribution reaches the boundary at $\phi=\phi_{\rm cr}$ and the
evolution of the system after that depends on the boundary conditions
imposed there.  There are two cases to consider.  If the energy flux
of the incoming matter is always smaller than the rate at which a
black
hole evaporates then the critical line remains time-like for
$x^+>x^+_0$ [\endpoint].  This requires the inequality
$$
T^f_{++}(x^+) < {\kappa\over 4{x^+}^2}
\eqn\fourteenth
$$
to hold for all values of $x^+$ in Kruskal coordinates.  If, on the
other hand, this inequality is violated at some value of $x^+$ then
the critical line will become space-like there and no meaningful
boundary conditions can be applied.  In this case an apparent horizon
will form to cloak the singularity and we have a black hole.  Once
the
incoming flux falls below the threshold value the apparent horizon
begins to recede and, as we discussed in our previous paper
[\endpoint], it will meet the singularity in a finite time proper
time.  When the evaporation of the black hole is complete the
critical line goes time-like again and we must impose boundary
conditions there.

For our discussion of boundary conditions we will assume that the
incoming energy flux remains below
threshold at all times.  This condition defines a low-energy sector
of the theory without real black holes (virtual black holes will
presumably appear at the quantum level) and it is of considerable
interest to study the quantum theory of this sector by itself.\foot{A
classical shock wave with
$T^f_{++}(x^+)={m\over \gl x^+_0}\delta(x^+-x^+_0)$
forms a black hole for arbitrarily small $m$.  This does not conflict
with the existence of a low-energy sector because such a concentrated
flux violates the uncertainty principle and is therefore not a good
description of a low energy state.}  For example, it would be very
interesting to determine whether quantum coherence loss occurs at a
non-vanishing rate in low-energy scattering as conjectured by Hawking
[\hawii].

We will derive our boundary conditions from the cosmic censorship
hypothesis as discussed in the introduction.  Imposing finite
curvature at the boundary is a coordinate invariant condition.  In
this section we work exclusively in Kruskal coordinates.  This allows
us to use special relations, such as $\Omega=\chi$, to simplify
calculations but at the same time
some of our formulas will appear non-covariant.

The curvature can be expressed in terms of
$\phi$ as follows,
$$\eqalign{
R =& -2 \nabla^2\rho   \cr
=& {4\over 1-{\kappa\over 4}e^{2\phi}} \,
\bigl[\gl^2 -(\nabla \phi)^2 \bigr] \>. \cr}
\eqn\curvature
$$
where we have used the semi-classical equations of motion.
The critical line is defined as the curve of constant $\phi$ where
$\Omega '(\phi)={\sqrt{\kappa}\over 2}
 ( 1-{\kappa\over 4}e^{2\phi}) = 0$
and the curvature
will only be finite at the boundary if the dilaton satisfies
$$
\nabla_n\phi = \gl
\eqn\dilbound
$$
there, where $\nabla_n$ denotes the invariant normal derivative at
the boundary.  Since $\Omega '(\phi)=0$ at the critical line it
follows that requiring
$$
\omp\big\vert_{\phi=\phi_{\rm cr}} =
\omm\big\vert_{\phi=\phi_{\rm cr}} = 0 \> ,
\eqn\sixteenth
$$
on the boundary where it is time-like is a necessary
condition for finite curvature there.  As we will see
below, this condition uniquely determines the solution in regions
(ii) and (iii) in figure 1.
At the end of this section we will show that
the resulting solution indeed has finite curvature on the boundary
which means that \sixteenth\ is also a sufficient
condition when we work in Kruskal coordinates.

Let us look for a solution in region~(ii) in figure~1 which matches
continuously onto the solution \thirteenth\ which holds in
region~(i).  The form of the new solution can only differ from
\thirteenth\ by some function of $x^-$ alone,
$$
\Omega^{(ii)}(x^+,x^-) =
\Omega^{(i)}(x^+,x^-) + F(x^-) \> ,
\eqn\seventeenth
$$
because otherwise the $++$ constraints would no longer be satisfied.
The superscripts on $\Omega$ refer to the region of validity in
figure~1.  The finite curvature conditions \sixteenth\ are sufficient
to determine both $F(x^-)$ and the shape of the boundary curve
$(\hxp,\hxm)$ in
terms of the incoming matter distribution.  For the solution
\seventeenth\ these conditions imply the following two relations,
$$\eqalign{
{\kappa\over 4} =&
-\gl^2 \hxp (\hxm +{1\over \gl^2}P_+(\hxp)) \> , \cr
\sqrt{\kappa} F '(\hxm) =&
\gl^2 \hxp + {\kappa \over 4\hxm} \> .  \cr}
\eqn\eighteenth
$$
The first one defines the critical line and the second one
can be integrated to obtain the matching function,
$$
F(x^-) =
{\sqrt{\kappa}\over 4} \log{(-\gl^2 x^-\hxp)}
- {M(\hxp)\over \sqrt{\kappa}\gl}
-{\sqrt{\kappa}\over 4} \log{({\kappa\over 4})}  \> .
\eqn\nineteenth
$$
In this way cosmic censorship determines a unique extension of the
solution into region~(ii).  There is no discontinuity in $\omm$ at
$x^-=x^-_0$ so
the $--$ constraints in \sixth\ are satisfied across the matching
line and there is no shock wave carrying energy out along this null
line.

By assumption the matter stops coming in at $x^+=x^+_1$ so $M(x^+)$
and $P_+(x^+)$ receive no contribution after that.  The solution
\seventeenth\ extends smoothly into region~(iii) where it takes the
form of a linear dilaton configuration with $x^-$ shifted by the
total incident Kruskal momentum,
$$
\Omega^{(iii)}(x^+,x^-) =
- {\gl^2\over \sqrt{\kappa}}\,
x^+\bigl(x^-+{1\over \gl^2}P_+(x^+_1)\bigr)
-{\sqrt{\kappa}\over 4}
\log{\bigl(-\gl^2 x^+(x^-+{1\over \gl^2}P_+(x^+_1))\bigr)} \> ,
\eqn\twentieth
$$
and there is no shock wave propagating out along $x^-=x^-_1$.

We will now give the boundary conditions implied by the cosmic
censorship relations \sixteenth\ a physical interpretation in terms
of reflected energy.  In Kruskal coordinates the constraints
\sixth\ reduce to
$$
0= \, \sqrt{\kappa}\, \p_\pm^2\chi
+ {1\over 2} \sum_{i=1}^N \p_\pm f_i \p_\pm f_i
-\kappa \, t_\pm \> .
\eqn\constr
$$
Let us evaluate $\p_\pm^2\chi$ at the boundary,
$$\eqalign{
\p_+^2\chi (\hat x) =&\,
- {1\over \sqrt{\kappa}} P_+'(\hxp)
+ {\sqrt{\kappa}\over 4(\hxp)^2} \> , \cr
\p_-^2\chi (\hat x) =&\,
F''(\hxm) + {\sqrt{\kappa}\over 4(\hxm)^2} \> . \cr}
\eqn\dtwo
$$
This expression holds everywhere if we define $F(x^-)=0$ in
region~(i).  Differentiating the first finite curvature relation
in \sixteenth\ with respect to $\hxp$ and the second one with
respect to $\hxm$ leads to the following relations,
$$\eqalign{
{\gl^2\over \sqrt{\kappa}} {d\hxm\over d\hxp} =&\,
- {1\over \sqrt{\kappa}} P_+'(\hxp)
+ {\sqrt{\kappa}\over 4(\hxp)^2} \> , \cr
{\gl^2\over \sqrt{\kappa}} {d\hxp\over d\hxm} =&\,
F''(\hxm) + {\sqrt{\kappa}\over 4(\hxm)^2} \> . \cr}
\eqn\onetwo
$$
By combining this with \dtwo\ and using the constraints we
obtain a reflection condition
$$
{\cal T}_{--} = ({d\hxp\over d\hxm})^2 \, {\cal T}_{++} \>
\eqn\reflect
$$
on the combination
$$
{\cal T}_{\pm\pm} =
{1\over 2} \sum_{i=1}^N \p_\pm f_i \p_\pm f_i
-\kappa t_\pm  \> .
\eqn\combo
$$

Let us assume that the physical incoming energy is in the form of
coherent radiation of matter fields.  We then expect the outgoing
radiation to consist of a coherent part and an incoherent one due
to the anomaly.  The reflection conditions \reflect\ obtained from
the cosmic censorship hypothesis do not separate the two
contributions and therefore they do not supply us with unambiguous
boundary conditions for the matter fields.  However, it appears
to be consistent with \reflect\ to impose reflecting boundary
conditions on the matter, for example
$$
f_i(\hat x) = 0\>.
\eqn\fref
$$
In this case the classical matter
energy momentum tensor by itself
would satisfy the reflection condition and it would follow from
\reflect\ that
$$
t_-(x^-) = ({d\hxp\over d\hxm})^2\, t_+\bigl(\hxp(x^-)\bigr)\>.
\eqn\tref
$$
As $x^-\rightarrow -\infty$ the semi-classical solution for
$\Omega$ should approach the corresponding classical solution
with the same incoming energy distribution.  This determines
$t_+={1\over 4 {x^+}^2}$ and for a given time-like boundary
curve \tref\ would give a unique $t_-$ describing the anomalous
component of the outgoing radiation.

Given this strong form of the reflecting conditions a distant
observer would be able to completely reconstruct the initial
state from the outgoing radiation and no information would be
lost in low-energy physics at the semi-classical level.  Quantum
fluctuations could still cause the boundary curve to go
space-like and lead to information loss.  It should be stressed
that while these strong reflection conditions, imposed directly on
the matter fields, are consistent with the cosmic censorship
hypothesis, they do not follow from it and it is possible that they
are not the appropriate boundary conditions at $\phi=\phi_{\rm cr}$.

Returning to the weaker form of the reflecting conditions
\reflect\ we can check their consistency by computing the
total energy radiated out to $x^+\rightarrow\infty$.
First of all, there should be no outgoing radiation in
region~(iii) in figure~1 where the solution is vacuum-like.
Evaluating ${\cal T}_{--}$ there gives
$$
{\cal T}_{--}^{(iii)} =
-{\kappa\over 4}{1\over (x^-+{1\over \gl^2}P_+(x^+_1))^2} \> .
\eqn\tthree
$$
This vacuum contribution must be subtracted from
${\cal T}_{--}$ to obtain the outgoing
energy in a given region and the total radiated energy is
$$
E_{out} = -\gl \int_{-\infty}^{x^-_1} dx^- \,
(x^-+{1\over \gl^2}P_+(x^+_1)) \,
\bigl[{\cal T}_{--}
+ {\kappa\over 4}
{1\over (x^-+{1\over \gl^2}P_+(x^+_1))^2} \bigr] \> .
\eqn\twentyeighth
$$
The weight factor of $-(x^-+{1\over \gl^2}P_+(x^+_1))$ appears
because we are not using asymptotically Minkowskian coordinates.  A
straightforward calculation shows that $E_{out}$ precisely equals the
total incoming energy $M(x^+_1)$ so our boundary conditions appear to
conserve energy.

Energy conservation can also be checked using a definition of total
energy in terms of the asymptotic curvature, analogous to the one
introduced in [\lslt],
$$
m(x^-)= \lim_{x^+\rightarrow\infty} {1\over 4\gl}
(1-{\kappa\over 4}e^{2\phi}) \,e^{-2\phi}\, R \> .
\eqn\twentyninth
$$
This definition gives the correct mass for classical black hole
solutions and tends to the total incoming energy as $x^-\rightarrow
-\infty$.  The formulas below are streamlined by including the factor
of $(1-{\kappa\over 4}e^{2\phi})$ in the definition of the energy.
It can be added at no cost since it goes to $1$ in the limit
$x^+\rightarrow\infty$.  Using the semi-classical equations of motion
for $\phi$ and $\rho$ the following expression is obtained for the
rate of change of the energy,
$$\eqalign{
{dm\over dx^-} =&
\lim_{x^+\rightarrow\infty} {2\over \gl} e^{-2\rho}\phip
(e^{-2\phi}-{\kappa\over 4})(2\p_-^2\phi-4\rhom\phim)  \cr
=&\lim_{x^+\rightarrow\infty} {2\over \gl} e^{-2\rho}\phip
\bigl[{\cal T}_{--} + \kappa(\rhom\rhom -\p_-^2\rho)\bigr] \>
,\cr}
\eqn\thirtieth
$$
where we used the constraints \third\ to obtain the second equality.
The term accompanying ${\cal T}_{--}$ inside the square brackets
subtracts off the same vacuum contribution as in \twentyeighth .  By
inserting the explicit semi-classical solution and integrating over
$x^-$ one can easily check that $m(x^-)$ goes to zero as we enter
region~(iii).

Since the conditions \sixteenth ,  which we used to determine the
solution in region~(ii), appear to be weaker than \dilbound , which
is expressed in terms of the original dilaton field, one might worry
that we have not guaranteed finite
curvature at the boundary.  However, the semi-classical geometry,
which results from imposing \sixteenth , is in fact non-singular as
the critical line is approached.
To see this, we consider a point near the boundary,
$(x^+,x^-)=(\hat x^+ + \delta x^+, \hat x^- + \delta x^-)$, and
evaluate the curvature in the limit of vanishing $\delta x^\pm$.
We need to show that the expression in the square brackets in
\curvature\
is of order $\delta x^\pm$.  In order to obtain
$\p_\pm\phi(\hat x)$ we expand $\Omega$ around
$\Omega(\phi_{\rm cr})$ in two different ways,
$$\eqalign{
\Omega(\phi) =&\, \Omega(\phi_{\rm cr})
+{1\over 2}\Omega ''(\phi_{\rm cr})(\phi -\phi_{\rm cr})^2
+\ldots  \cr
=&\, {\sqrt{\kappa}\over 4}(1-\log{\kappa\over 4})
+{\sqrt{\kappa}\over 2}
(\phi -\phi_{\rm cr})^2 +\ldots\> ,\cr}
\eqn\expone
$$
and
$$\Omega(x^+,x^-)=\Omega(\hat x)
+{1\over 2}\p_+^2\Omega(\hat x){\delta x^+}^2
+\p_+\p_-\Omega(\hat x)\delta x^+\delta x^-
+{1\over 2}\p_-^2\Omega(\hat x){\delta x^-}^2
+\ldots \>.
\eqn\exptwo
$$
Comparing \dtwo\ and \onetwo\ leads us to write \
$\p_+^2\Omega(x)={\gl^2\over\sqrt{\kappa}}\,
{d\hxm\over d\hxp}$
and
$\p_-^2\Omega(x)={\gl^2\over\sqrt{\kappa}}\,
{d\hxp\over d\hxm}$, where we have used that $\Omega=\chi$ in Kruskal
coordinates. Inserting these relations, along with the
equation of motion, $\partial_+\partial_-\Omega =
-{\gl^2\over \sqrt{\kappa}}$, into \exptwo\ and comparing with
\expone\ gives the following expression for $\phi$ in terms of
$\delta x^\pm$,
$$
\phi -\phi_{\rm cr} = {\gl\over \sqrt{\kappa}}
\bigl[ \sqrt{d\hxp\over d\hxm}\, \delta x^-
-\sqrt{d\hxm\over d\hxp}\, \delta x^+ \bigr] +\ldots \>.
\eqn\expthree
$$
{}From this we can read off the values of
$\phip (\hat x)$ and $\phim (\hat x)$ and insert them into
\curvature\ to see that our solution has finite curvature at
$\phi=\phi_{\rm cr}$.

Finally it is interesting to note that the solution for
$\Omega$ in region (ii) can be expressed in terms of the boundary curve in a
very simple manner,
$$\eqalign{
\Omega(x^+,x^-) -\Omega_{\rm cr}
=&\, {\gl^2\over \sqrt{\kappa}} \bigl[
-x^+x^- + \hxp(x^-)x^-
+\int^{x^+}_{\hxp(x^-)}du^+\,\hxm(u^+)
\bigr] \cr
=&\, {\gl^2\over \sqrt{\kappa}} \bigl[
-x^+x^- + x^+\hxm(x^+)
+\int^{x^-}_{\hxm(x^+)}du^-\,\hxp(u^-)
\bigr]\> . \cr}
\eqn\thirtyfirst
$$
These relations can be obtained by integrating the expressions
for $\p_+^2\Omega(x)$ and $\p_-^2\Omega(x)$ given above.

\chapter{Discussion}

We have shown how the cosmic censorship hypothesis leads to
reflecting boundary conditions for matter energy.  The matter
carrying this reflected energy consists of both coherent and
incoherent $f$-fields and cosmic censorship alone does not
allow us to distinguish between the two components.  It is
nevertheless strong enough to uniquely determine the
semi-classical evolution of the geometry and dilaton field
for a given distribution of low-energy incoming matter and
we have checked that energy is conserved in that evolution.
Other boundary conditions which do not rule out extended naked
singularities presumably lead to instability.
They could allow a black hole to evolve into an object
carrying an arbitrary amount of negative energy
and it could then continue to radiate forever happily.
If we implement boundary conditions consistent with cosmic
censorship an evaporating black hole returns to the vacuum
configuration after a finite proper time and the Hawking
emission stops [\endpoint].

The expression for the curvature \curvature\ takes a simple form on
the apparent horizon of a black hole,
$$
R\Big\vert_{\phip=0} =
{4\gl^2 \over  1 - {\kappa\over 4}e^{2\phi}}  \> .
\eqn\horcurv
$$
{}From this it is clear that the curvature on the apparent horizon
diverges as it approaches the singularity curve and therefore cosmic
censorship is violated at the endpoint of the evaporation process.
If the boundary conditions we have advocated here are adopted the
violation of cosmic censorship is minimal, in that the naked
singularity is an isolated event.

At the semi-classical level it appears that information loss can
be avoided in the low-energy sector of the theory by imposing
reflecting boundary conditions directly on the matter fields.
This is a desirable feature of any model of real low-energy
physics.  If, on the other hand, the incoming energy flux is
above the threshold for forming a black hole we see no way to
recover any information about the initial state, except its total
energy, in this semi-classical theory.  It remains an interesting
open question whether quantum fluctuations cause the boundary
to go space-like even in the low-energy sector and whether this
leads to disastrous loss of quantum coherence [\bps].

\FIG\fyrsta{Kruskal diagram of a semi-classical geometry with
low-energy matter incident on the linear dilaton vacuum.  The thick
line is the space-time boundary at $\phi=\phi_{\rm cr}$.}

\refout
\figout
\end